\documentclass[a4paper, oneside, twocolumn, notitlepage, 10pt]{extarticle_ecoc}
\usepackage{ecoc}
\usepackage{lineno}
\usepackage{algorithm}
\usepackage{algpseudocode}
\usepackage{mathptmx} % Use Times Font
\usepackage{cleveref}
\usepackage{subcaption}
\usepackage{setspace}
\usepackage{todonotes}
\usepackage{verbatim}
\usepackage{siunitx}
\DeclareSIUnit{\belmilliwatt}{Bm}
\DeclareSIUnit{\dBm}{\deci\belmilliwatt}
\DeclareSIUnit{\sample}{Sa}

\begin{document}
\selectlanguage{english}    % Standard Language

%-------------------------------------------------- Title -----------------------------------------------------%

 \title{Quantum Noise Limited Temperature-Change Estimation for $\Phi$-OTDR Employing Coherent Detection}
 
% \title{Approaching the Quantum Noise Limited Phase Estimation for Phase-Coherent OTDR}%
% \title{Quantifying Temperature Change Uncertainty in Phase-Coherent OTDR: From Quantum Noise to Practical Sensing Limits}%
% \title{Quantum Noise-Limited Temperature Resolution in Phase-Coherent OTDR}%
% \title{Quantum Noise Limited Temperature Rate Estimation and Uncertainty Analysis in Phase-Coherent OTDR}%
% \title{From Phase Noise to Temperature Precision: Fundamental Limits in Coherent Φ-OTDR Sensing}%
% \title{Approaching Quantum Limits: Phase and Temperature Estimation in Distributed Coherent OTDR}%
% \title{Quantum Noise Limited Phase and Temperature Estimation in Phase-Coherent OTDR Systems}%
% \title{From Phase Estimation to Temperature Sensing: Fundamental Limits and Uncertainty Analysis in Phase-Coherent OTDR}%
%\title{From Phase Estimation to Temperature Sensing: Approaching the Quantum Noise Limited Performance in Phase-Coherent OTDR}%

%------------------------------------------------- Authors-----------------------------------------------------%

\author{
    Huwei Wang\textsuperscript{(1)}, Roman Ermakov\textsuperscript{(1)},
    Francesco Da Ros\textsuperscript{(1)}, Darko Zibar\textsuperscript{(1)}
}

\maketitle                  % Create title and author

%------------------------------------------ Description of Authors ----------------------------------------------%

\begin{strip}
    \begin{author_descr}

        \textsuperscript{(1)} Department of Electrical and Photonics Engineering, Technical University of Denmark, Kongens Lyngby 2800, Denmark,
        \textcolor{blue}{\uline{huwwa@dtu.dk} }

        % \textsuperscript{(2)} Authors full affiliation,
        % \textcolor{blue}{\uline{author@institution.org}} (Email address optional)

        % \textsuperscript{(3)} Authors full affiliation,
        % \textcolor{blue}{\uline{author@institution.org} (Email address optional)

    \end{author_descr}
\end{strip}

% \setstretch{1.1}
%-------------------------------------------------- Footnote -------------------------------------------------------%
\renewcommand\footnotemark{}
\renewcommand\footnoterule{}
%\let\thefootnote\relax\footnotetext{text}

%-------------------------------------------------- Abstract ---------------------------------------------------------%

\begin{strip}
    \begin{ecoc_abstract}
    The quantum limit is a fundamental lower bound on the uncertainty when estimating a parameter in a system dominated by the minimum amount of noise (quantum noise). For the first time, we derive and demonstrate a quantum limit for temperature-change estimation for coherent phase-OTDR sensing-systems.  \textcopyright2025 The Author(s)
    \end{ecoc_abstract}
\end{strip}

% We investigate the quantum noise limit of phase and temperature change estimation in phase-coherent OTDR systems. Through large-scale simulation, we demonstrate phase estimation approaches the quantum limit and explicitly quantify how phase uncertainty fundamentally constrains temperature measurement precision.
% This study investigated the sensing phase fluctuation in phase-coherent OTDR systems by simulating Rayleigh backscattering. Results demonstrate logarithmic phase-fluctuation reduction with SNR, aligning closely with theoretical quantum limits. We also derived the phase change expression with temperature rate and further revealed the explicit expression for the estimated temperature change uncertainty.
%-------------------------------------------------- Introduction Section -------------------------------------------------------%

\section{Introduction}
% Optical fiber sensing technology plays a crucial role in the monitoring of critical infrastructure, including fiber-optic communication networks, power grids, water distribution systems, and natural gas production pipelines \cite{marie2021principle,peng2014ultra}. Among these technologies, sensing systems based on phase-sensitive optical time domain reflectometry ($\Phi$-OTDR), particularly when integrated with digital coherent detection, are gaining significant attention. This is due to their capability to deliver long-range, high-resolution monitoring of key parameters such as temperature, strain, and vibrations \cite{shaheen2023phase,Zhong:21,soriano2023time}.

% Compared to conventional non-coherent OTDR systems, digital coherent detection enables linear detection of the Rayleigh backscattered optical field, allowing for application of digital signla processing technqiues for optimal optical phase estimation \cite{liehr2018wavelength}. Accurate phase estimation is critical, as variations in temperature, strain, and vibration induce changes of the optical path length in the optical fiber, which in turn manifest as phase shifts in the Rayleigh backscattered signal. Any uncertainty in the estimation of this optical phase directly contributes to uncertainty in the measurement of temperature, strain, and vibration. Therefore, minimizing phase estimation error is key to enhancing the overall accuracy and reliability of $\Phi$-OTDR-based sensing systems \cite{lu2022phase,wu2024phase}.

Phase-sensitive optical time-domain reflectometry ($\Phi$-OTDR) plays a vital role in monitoring critical infrastructure such as fiber-optic communication networks, power grids, water distribution systems and natural gas pipelines \cite{marie2021principle,shaheen2023phase,soriano2023time}. Compared to traditional non-coherent OTDR systems, digital coherent detection enables linear reconstruction of the Rayleigh backscattered optical field in the digital domain, allowing for more precise estimation of key physical monitoring parameters (temperature, strain etc)\cite{liehr2018wavelength,lu2010distributed}.

Extensive research has been conducted on coherent $\Phi$-OTDR for temperature-change monitoring \cite{marcon2019analysis,lu2019numerical,liu2018true}. To the best of our knowledge, the fundamental limits of the temperature-change estimation accuracy have not been thoroughly investigated. In general, the accuracy in estimating any physical parameter is influenced by the level of noise in the system. The maximum achievable accuracy (lowest achievable uncertainty) is obtained when the system is dominated by the minimum amount of noise - quantum noise.

In coherently detected $\Phi$-OTDR fiber sensing systems, the primary noise sources include optical amplifier noise, as well as thermal and shot noise intrinsic to the coherent detection process. When the system operates in the quantum-noise-limited regime, the system has a minimum amount of noise and the signal-to-noise ratio (SNR) is thereby maximized. 
% The SNR is maximized in the shot-noise-limited regime, where quantum noise becomes the dominant factor. 
In this regime, we investigate the fundamental quantum limit on temperature-change estimation defining the ultimate achievable precision and analyzing its scaling behavior with respect to the SNR and the monitoring length.

Changes in the temperature  will induce changes of the optical path length in the sensing fiber. This will manifest as phase shifts in the coherently detected Rayleigh backscattered signal. The estimation of the temperature changes can then be computed from the corresponding phase shifts \cite{martins2013coherent}.  Accurate optical phase estimation is critical as any uncertainty in the estimation of this optical phase directly contributes to uncertainty in the estimation of the temperature-change. To minimize the uncertainty in the temperature change estimation, the optical phase estimation must therefore be performed at the quantum limit. 

Extensive research has been conducted on approaching the quantum limit of optical phase estimation for shot-noise limited coherently detection \cite{zibar2021approaching,berry2002adaptive}. Compared to \cite{zibar2021approaching}, coherently detected $\Phi$-OTDR systems contain a sensing fiber that provides the back-scattered Rayleigh signal. 
The interaction between the optical amplifier noise, the back-scattered Rayleigh signal and receiver noise may exhibit complex behavior. 
%The stochasticity of amplifier noise, Rayleigh scattering and receiver noise leads to a complex behavior. 
In this paper, we numerically investigate the quantum noise limited phase estimation for back-scattered Rayleigh signal. For our simulations, we employ a highly-accurate models of the Rayleigh back-scattering\cite{liokumovich2015fundamentals}. We show that, for $\Phi$-OTDR, the quantum limit for the optical phase estimation approaches the quantum limit of the shot-noise limited coherent detection. Using this results, we derive a quantum limit for how accurately temperature-change can be estimated. 

\begin{figure*}[htb]
\centering
\includegraphics[width=\textwidth]{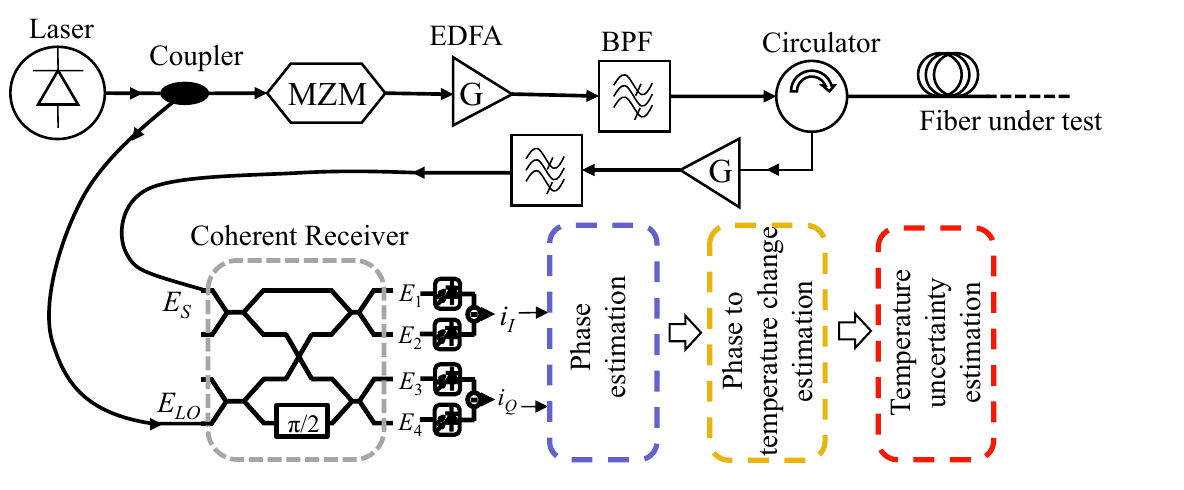}
\caption{Simulation set-up for coherently detected $\Phi$-OTDR system.} 
% MZM: Mach-Zehnder Modulator, EDFA: Erbium-Doped Fiber Amplifier, BPF: Band Pass Filter, ADC: Analog-to-Digital Converter
\label{fig:Systemsetup}
% \rule{\textwidth}{0.5pt} % 添加横线
\end{figure*}

\begin{figure}[t]
\centering
\includegraphics[width=1.05\linewidth]{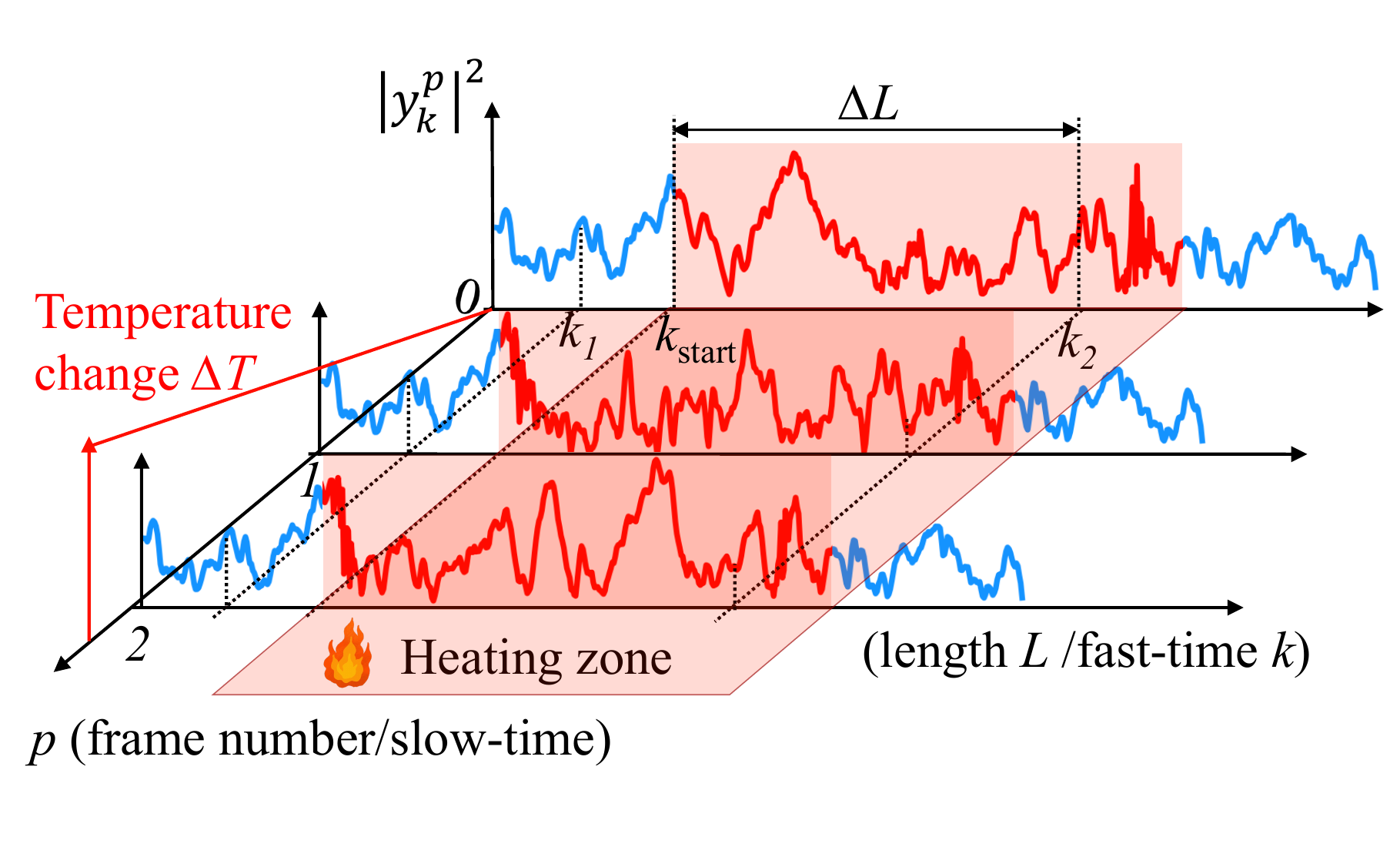}
\caption{Backscattered signal in slow time domain and fast time domain.}
\label{fig:slowfast}
\end{figure}

\section{System Setup}
The numerical setup used to simulate a coherently detected $\Phi$-OTDR fiber-based sensing system is illustrated in Fig. \ref{fig:Systemsetup}. The system comprises a continuous wave (CW) laser source, a pulse generator, a Mach-Zehnder modulator (MZM), two erbium-doped fiber amplifiers (EDFA), two bandpass filters (BPF), a circulator, the fiber under test, and a coherent receiver. The CW laser emits light at a wavelength of $\lambda_0$=\qty{1550}{\nm} and is then split into a signal and local oscillator (LO). The signal part of the CW is shaped into rectangular optical pulses $r(t)$, using a MZM. The pulse repetition rate and pulse duration are set to $f_P=$ \qty{100}{\kilo\hertz} and $\tau=$ \qty{100}{\ns}, respectively. The modulated optical signal is then amplified in an EDFA with a gain of \qty{22}{\dB} and a minimum noise figure (NF) of \qty{3}{\dB}. The peak power of the optical signal entering the fiber is \qty{23}{\dBm}.   

 The sensing fiber is a \qty{40}{\km} long single-mode fiber with an attenuation coefficient $\alpha_{att}$ of \qty{0.2}{\dB\per\kilo\metre}. The Rayleigh backscattered signal is returned through the third port of the circulator and it is amplified with an EDFA with a gain of \qty{22}{\dB} and NF of \qty{3}{\dB}. For the simulation of the back-reflected Rayleigh signal we implemented the model reported in\cite{liokumovich2015fundamentals}.  
 
 The amplified Rayleigh backscattered signal is then mixed with a local oscillator (LO), with a power of \qty{0}{\dBm}, in a coherent receiver with a bandwidth of $B=$ \qty{300}{\mega\hertz}. The coherent receiver detects in-phase and quadrature components of back-scattered Rayleigh signal and is assumed to operate in the shot-noise limited mode. After detection the backscattered signal is digitized by an analogue to digital converter (ADC) with a sampling frequency of $F_s=$ \qty{625}{\mega\sample\per\second}. The backs-cattered signal is then reconstructed as: $y^p_k = i^p_k + jq^p_k$. We simulate one pulse at a time.  

\section{Phase and temperate-change estimation}
Let integer $k = 1:K$ be a discrete time index representing fast time, where $k \equiv kT_s$, and $T_s = 1/F_s$. The total number of fast-time samples for a single backscattered pulse is given by $K = \lfloor {\frac{2L_{\text{FUT}} F_s}{v_g}} \rfloor$, where $L_{\text{FUT}}=$\qty{40}{\km} is the length of the fiber under test (FUT), and $v_g$ is the group velocity of light in the fiber. The sensing length corresponding to index $k$ is thus $L_k = {\frac{k v_g}{2F_s}}$.

Let integer $p = 1:P$ be a discrete time index representing slow time, indexing each back-reflected Rayleigh signal associated with the $p$-th probe pulse. An example of the magnitude squared of the back-reflected signal, $|y_k^p|^2$, is shown in Fig.~\ref{fig:slowfast} for $p = 0,1,2$. The figure also illustrates a gradual temperature increase $\Delta T$ over a finite segment of the fiber. Although the round-trip time depends on the FUT length $L_{\text{FUT}}$, it typically does not exceed $\sim\!1\,\mathrm{ms}$. Therefore, the slow process of temperature variation can be assumed to occur at a fixed rate over many frames in the heating zone.

As shown in Fig.~\ref{fig:slowfast}, the back-reflected signal $y_k^p$ is passed to a phase estimation block, where the phase is estimated using $\hat{\phi}_k^p = \tan^{-1}(y_k^p)$. The estimated phase $\hat{\phi}_k^p$ is analyzed over slow time (index $p$) to identify a reference location $k_1$, at a distance $L_1 = k_1 v_g T_s/2$, which is unaffected by temperature changes. In contrast, a second location $k_2$, at $L_2 = k_2 v_g T_s/2$, is influenced by the temperature affection and exhibits a linear phase change with respect to $p$, compared to the stable reference $k_1$.

To monitor temperature changes, the phase difference between the two sensing points is computed per frame $p$ as
$\Delta \hat{\phi}^{k_2-k_1}(p) = \hat{\phi}_{k_2}^p - \hat{\phi}_{k_1}^p$. We then apply the method introduced in our recent work~\cite{ermakov2025method} to convert the estimated phase difference $\Delta \hat{\phi}^{k_2-k_1}(p)$ into a temperature change estimate $\Delta \hat{T}^{k_2}(p)$ at position $k_2$.

The next objective is to evaluate the uncertainty (standard deviation) $\sigma_{\Delta T}$ of the temperature-change estimate $\Delta \hat{T}^{k_2}(p)$, which is derived from the uncertainty $\sigma_{\Delta \phi}$ of the phase change estimate $\Delta \hat{\phi}^{k_2-k_1}(p)$. Assuming minimal system noise (quantum-limited performance), $\sigma_{\Delta \phi}$ and $\sigma_{\Delta T}$ represent the fundamental limits of phase- and temperature-change estimation accuracy.

\begin{figure}[t]
\centering
\includegraphics[width=\linewidth]{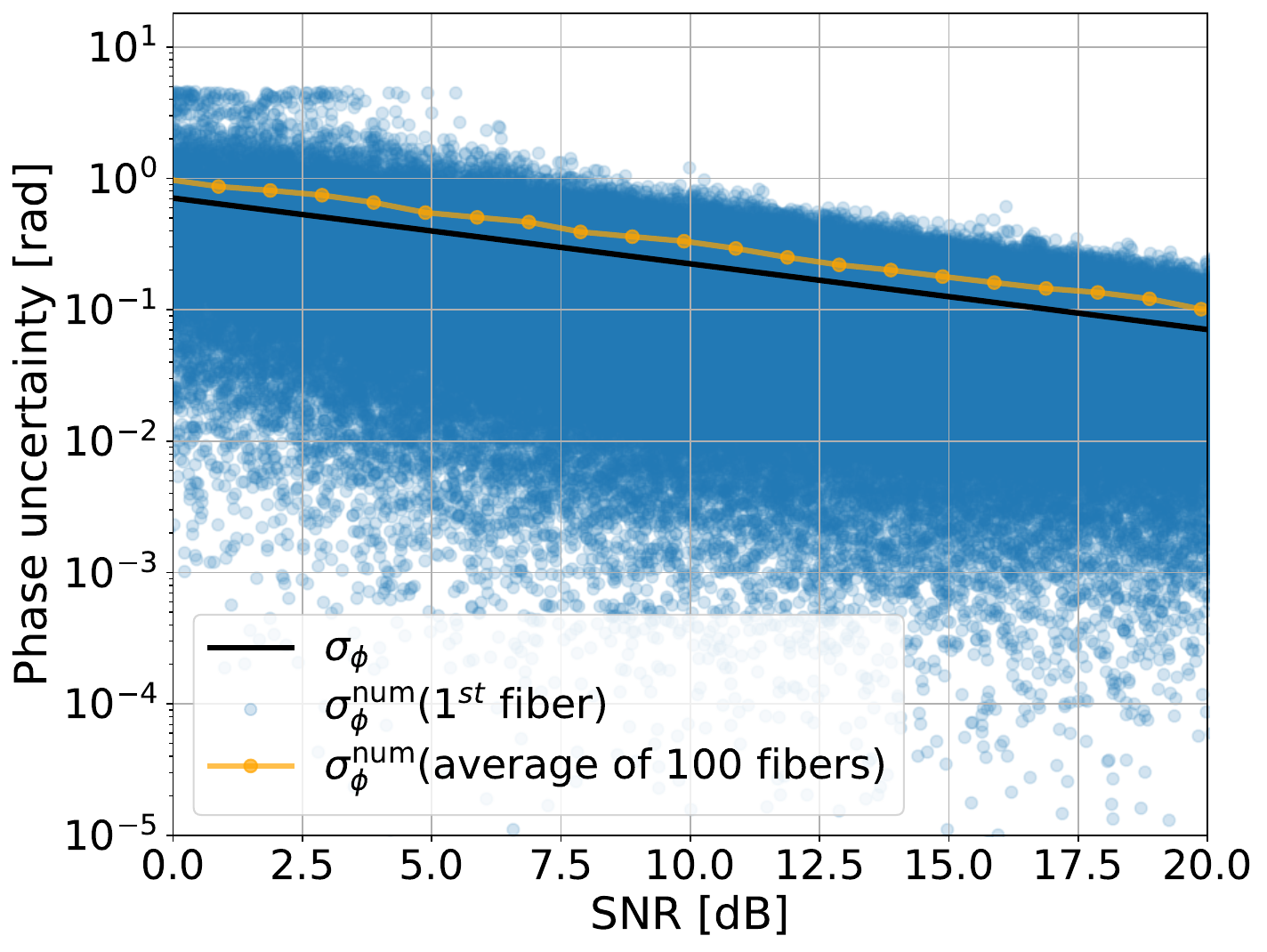}
\caption{Phase uncertainty as a function of SNR}
\label{fig:PhaseFluc}
% \rule{\linewidth}{0.5pt}
\end{figure}
% \vspace{0.5cm}

Due to space constraints, we omit the full derivation and present only the final result. A key assumption in the derivation is that the estimated phase samples of the Rayleigh backscattered signal are uncorrelated, which holds true when optical amplifier noise and receiver shot noise dominate the system noise, as supported in~\cite{vidal2023noise}.

We derive the uncertainty on the estimation of temperature-change from frame, $p$, to frame, $p+1$, to be: 
 \begin{equation}
\label{eq:QL DT}
\begin{split}
\sigma^{p\rightarrow p+1}_{\Delta T}  = \frac{\sqrt{2} }{C_f \Delta L} \sigma_ {\phi}
\end{split}
\end{equation}
where $\sigma_{\phi}$ is the minimum uncertainty on the phase estimation along the fast time. $C_f$ is a constant that depends on the laser source and sensing fiber parameters. $\Delta L = L_{{2}}-L_{{start}}$ is the length between the monitoring point (inside the heating zone) and the start of the heating zone, as shown in Fig.~\ref{fig:slowfast}.

\section{Results}
Computing the uncertainty on the estimation of temperature-change in Eq.~(\ref{eq:QL DT}), requires computing computing the uncertainty in phase $\sigma_ {\phi}$ along the fast time. It the absence of the sensing fiber, following the approach presented in\cite{zibar2021approaching}, 
the analytical phase uncertainty $\sigma_{\phi}$ can expressed as:    

\begin{equation}
\label{eq:sigma_coh}
\sigma_{\phi} =\sqrt {\frac{1}{\mathrm{2 SNR}}}
\end{equation}
where $\mathrm{SNR}$ is the signal-to-noise ratio of the signal after coherent detection computed in a bandwidth $B$. Next, we perform numerical simulation  including the sensing fiber and use the aforementioned procedure to extract the corresponding phase uncertainty as a function of $\mathrm{SNR}$. To compute the standard deviation (uncertainty), $\sigma^{num}_{\phi}$, the estimated phase needs to be compared to the true phase, which in our case is the extracted phase in the absence of any noise. In order to get sufficient statistics, we average over 100 different fibers. In Fig.~\ref{fig:PhaseFluc}, the numerically computed phase uncertainty is plotted as a function of $\mathrm{SNR}$. As a reference, we also plot Eq.~(\ref{eq:sigma_coh}). It is observed that that there is only a slight gap between $\sigma^{num}_{\phi}$ and $\sigma_{\phi}$. This implies that the Rayleigh backscattering introduces only a very limited uncertainty in the estimation. The difference may be attributed to the fact that,  the actual simulation inevitably involves a finite number of scatterers within each resolution cell, leading to slight deviations from the ideal Rayleigh distribution and introducing additional phase fluctuations.
% while the theoretical model assumes ideal complex Gaussian statistics for the Rayleigh backscattered signal,

\begin{figure}[t]
\centering
\includegraphics[width=\linewidth]{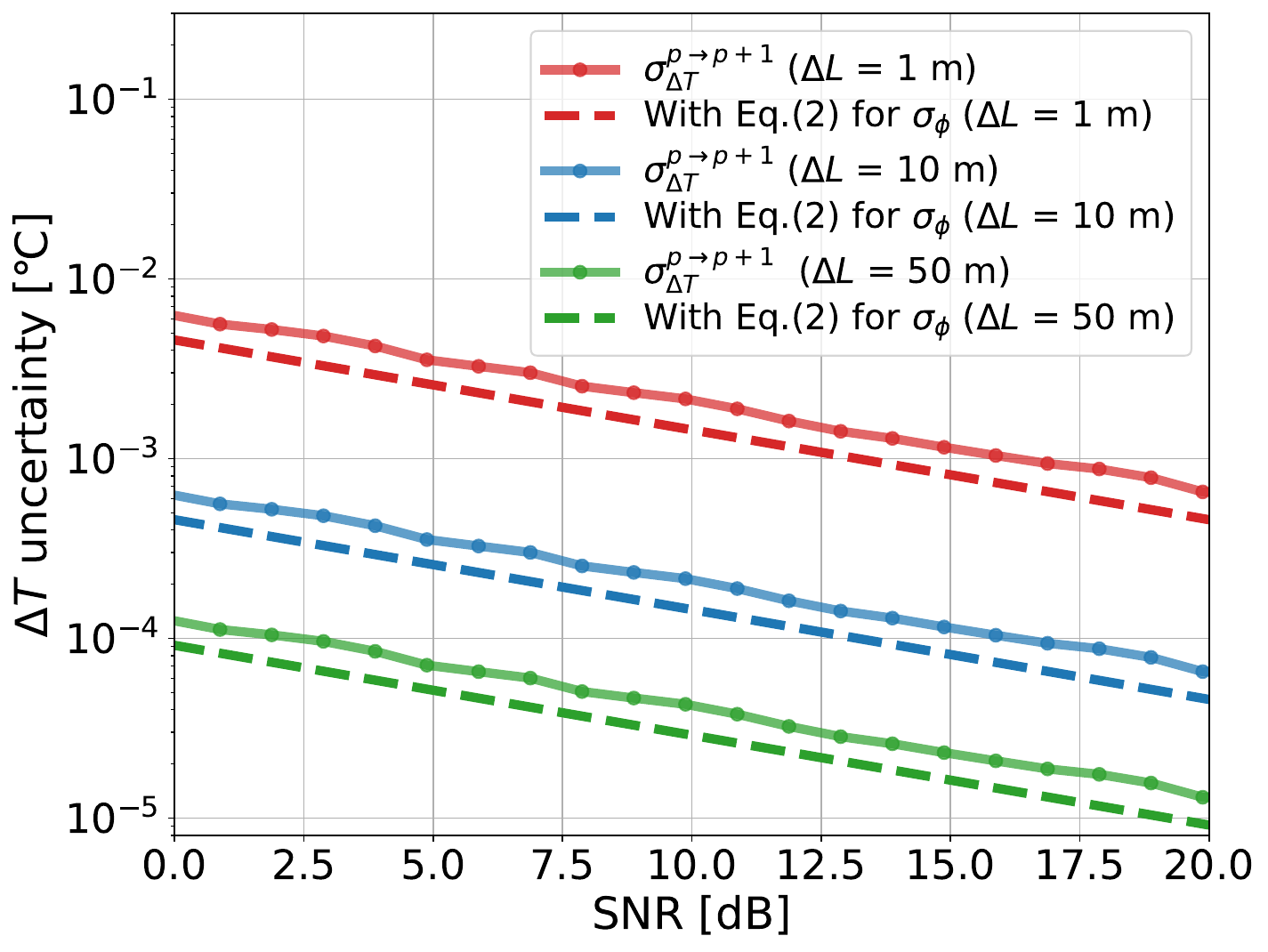}
\caption{Estimated temperature-change uncertainty as a function of SNR and for different sensing lengths $\Delta L$}
\label{fig:TempUncer}
% \rule{\linewidth}{0.5pt}
\end{figure}

In order to move from phase to temperature uncertainty, we use Eq.~(\ref{eq:QL DT}) to compute the temperature change uncertainty, when moving from one frame to another, for different sensing lengths $\Delta L$. We use the numerically obtained phase uncertainty, $\sigma^{num}_{\phi}$ as well as the one from Eq.~(\ref{eq:sigma_coh}). The results are plotted in Fig.~(\ref{fig:TempUncer}). The sensing length, $\Delta L$ is varied from 1 m to 50 m. The uncertainty in temperature-change estimation increases as the sensing distance is decreased. In general, it can be observed that within the considered ranges in Fig.~\ref{fig:TempUncer}, the uncertainty is relatively small, which is a promising results.
As the number of frames increases, the temperature change uncertainty would decrease with the number of frames due to the averaging of random fluctuations.

\section{Conclusions}
We have derived and numerically demonstrated a quantum limit on the uncertainty in estimating temperature changes in $\Phi-$OTDR systems. Our results reveal a linear relationship between the uncertainty in phase estimation and that of temperature change. This quantum limit serves as a benchmark for evaluating how accurately temperature changes can be estimated for $\Phi-$OTDR sensing systems, and how this accuracy scales with the sensing length.

\clearpage
\section{Acknowledgements}
This work is funded by the European Commission Horizon Europe Framework Program under SoFiN Project (Grant No. 101093015) and Villum Fonden (VIL5448, VIL29344).

%-------------------------------------------------- Bibliography Section -------------------------------------------------------%
% see also https://tex.stackexchange.com/questions/55030/text-before-references-but-after-bibliography-title-with-bibtex as of 2024-02-29
\defbibnote{myprenote}{%
% Citations must be easy and quick to find. More precisely:
% \begin{itemize}
%     \item Please list all the authors. 
%     \item The title must be given in full length. 
%     \item Journal and conference names should not be abbreviated but rather given in full length.
%     \item The DOI number should be added incl. a link.
% \end{itemize}
}
\printbibliography[prenote=myprenote]

\vspace{-4mm}

%%%%%%%%%%%%%%%%%%%%%%%%%%%%%%%%%%%%%%%%%%%%%
%---------------------------------------------- End of Document -----------------------------------------------%
\end{document}